\documentclass[12pt]{article}
\usepackage{amssymb}
\usepackage[dvipdfmx]{graphicx}
\usepackage{indentfirst}

\begin{document}

\begin{titlepage}
\begin{center}

\vspace*{25mm}

{\LARGE\bf On D-brane dynamics and moduli stabilization}

\vspace*{25mm}

{\large 
Noriaki Kitazawa
}
\vspace{10mm}

Department of Physics, Tokyo Metropolitan University,\\
Hachioji, Tokyo 192-0397, Japan\\
e-mail: noriaki.kitazawa@tmu.ac.jp

\vspace*{25mm}

\begin{abstract}
We discuss the effect of the dynamics of D-branes on moduli stabilization
 in type IIB string theory compactifications, with reference to a concrete toy model of
 $T^6/{\bf Z}_3$ orientifold compactification with fractional D3-branes and anti-D3-branes
 at orbifold fixed points.
The resulting attractive forces between anti-D3-branes and D3-branes,
 together with the repulsive forces between anti-D3-branes and O3-planes,
 can affect the stability of the compact space.
There are no complex structure moduli in $T^6/{\bf Z}_3$ orientifold,
 which should thus capture some generic features of more general settings
 where all complex structure moduli are stabilized by three-form fluxes.
The simultaneous presence of branes and anti-branes brings along the breaking of supersymmetry.
Non-BPS combinations of this type are typical of ``brane supersymmetry breaking'',
 and are a necessary ingredient in the KKLT scenario for stabilizing the remaining K\"ahler moduli.
The conclusion of our analysis is that,
 while mutual D-brane interactions sometimes help K\"ahler moduli stabilization,
 this is not always the case.
\end{abstract}

\end{center}
\end{titlepage}

\section{Introduction}
\label{introduction}

Orientifold models~\cite{orientifolds}, or string models with D-branes,
 provide a wide class of realizations of String Theory
 that can be useful tools to approach low--energy physics,
 on the par with corresponding models of Quantum Field Theory,
 for which they are expected to provide ultraviolet completions.
D-branes in type II superstring theories~\cite{polchinski} are their main new ingredients:
 they bring along gauge symmetries and matter fields,
 whose interactions reflect their configurations in the six-dimensional compact space
 (for reviews of these constructions see also~\cite{Blumenhagen:2006ci,Ibanez:2012zz},
  and references therein).
D-branes have tensions
 which result in large positive contributions to the four--dimensional cosmological constant,
 so that flat vacua require the presence of other objects with negative tension.
This task can be performed by the standard O${}_-$ orientifold planes,
 which are BPS objects carrying negative tensions and can compensate D-brane charges.
This is the setting that we shall refer to here,
 albeit allowing for the simultaneous presence of branes and anti-branes
 in order to break supersymmetry,
 although these can give rise to tachyon instabilities when they come too close to one another.
O${}_+$ planes, which carry positive tension and charge,
 provide a tachyon--free alternative,
 which is usually referred to as ``brane supersymmetry breaking''~\cite{bsb},
 but leaving aside the tachyon issue the resulting low--energy descriptions have some similarities. 
Furthermore, the so--called KKLT uplift~\cite{KKLT}
 utilizes the residual brane (or brane-orientifold) tensions,
 and it is identical in structure to the ubiquitous signature of brane supersymmetry breaking.

Supersymmetry breaking is a deep and non--trivial phenomenon
 that ought to accompany other non--trivial phenomena
 like spontaneous gauge symmetry breaking with a stabilized order parameter field,
 or the stabilization of moduli of six-dimensional compact space
 within the KKLT scenario~\cite{KKLT}, for example.
As we have anticipated,
 here we shall concentrate on its introduction
 by the simultaneous presence of D-branes and anti-D-branes,
 subject to the condition that Ramond--Ramond tadpole cancelation holds.
The presence of anti-D-branes has non-trivial effects
 on the stabilization of moduli associated to the six-dimensional compact space,
 especially when the D-branes are placed at singularities~\cite{Aldazabal:2000sa,Kitazawa:2014hya}.
If some D-branes and anti-D-branes are fractional branes,
 namely, they are fixed at the corresponding singularities,
 shorter distances between these fixed points are preferred
 because of the resulting attractive forces.
The same happens for the combinations of anti-D-branes and O-planes
 in brane supersymmetry breaking,
 because of the net attractive force between them.
All these types of D-brane dynamics
 affects moduli stabilization scenarios~\cite{KKLT,Balasubramanian:2005zx,Antoniadis:2004pp},
 and can favor it or disturb it, depending on the detailed nature of the configurations.

This article is devoted
 to illustrating the non-trivial effects of D-brane dynamics on moduli stabilization
 with reference to a concrete toy model of a $T^6/{\bf Z}_3$ orientifold compactification,
 where differently from the supersymmetric case described in~\cite{Angelantonj:1996uy}
 we allow for the simultaneous presence of fractional D-branes and anti-D-branes.
Our aim is not to construct a realistic model of moduli stabilization with D-brane dynamics,
 but more modestly to illustrate and discuss possible contributions to the potential of moduli
 from D-brane dynamics.

In the next section we begin by introducing the model $T^6/{\bf Z}_3$ orientifold
 that we plan to analyze.
The orientifold projection is effected by the operator $(-1)^{F_L} R_1 R_2 R_3 \Omega$,
 where $R_i$ is the reflection of $i$-th factorized torus in $T^6 = T^2 \times T^2 \times T^2$
 and $F_L$ is the left-handed space-time fermion number.
This orientifold projection introduces O3-planes at each orientifold fixed point.
The model comprises
 56 fractional D3-branes and 24 fractional anti-D3-branes at orbifold fixed points,
 together with 64 O3-planes at orientifold fixed points in compact $T^6/{\bf Z}_3$ space.
In order to cancel twisted Ramond--Ramond tadpoles at each fixed point,
 12 fractional D7-branes and 12 fractional anti-D7-branes at orbifold fixed points
 are also introduced.
Untwisted Ramond--Ramond tadpoles are also canceled,
 and the total Ramond--Ramond charges vanish in the compact $T^6/{\bf Z}_3$ space.
The positions of these objects are explicitly described by vectors in compact space
 using the technique of Coxeter orbifolds~\cite{Kobayashi:1991rp},
 and 9 moduli parameters are identified in inner products of six lattice basis vectors.
In section \ref{potential} we derive explicitly
 the potential energies that result from the attraction between anti-D3-branes and D3-branes,
 from the repulsion between anti-D3-branes and O3-planes
 and from the attraction between anti-D7-branes and D7-branes.
We do not account for the torus periodicity,
 so that the potential takes relatively simple forms
 in terms of the shortest distances between two objects.
This simplification,
 without concentrating too much about special features of a toy model,
 is convenient to qualitatively understand possible universal effects of the D-brane dynamics
 on moduli stabilization.
After a simple preliminary discussion based on an ansatz on moduli parameters,
 we also provide a numerical analyses of a general case.
As we shall seem three volume moduli are not stabilized by the D-brane dynamics only,
 because one can not balance exactly, in general,
 the attractive forces between anti-D3-branes and D3-branes
 and the repulsive forces between anti-D3-branes and O3-planes.
As a result,
 the volume moduli stabilization by KKLT scenario, for example,
 should take this kind of unstable potential as a starting point,
 rather than the original flat potential.
In detail, we shall see that four of the original moduli can be stabilized by the D-brane dynamics,
 if three volume moduli can be stabilized by some other mechanism.
Moreover, one of the volume moduli can be stabilized
 by the additional attractive forces between anti-D7-branes and D7-branes,
 if the two other volume moduli are stabilized by some other mechanism.
Section \ref{conclusions} contains a summary and discussion.

\section{The model}
\label{model}

Consider a six-dimensional $T^6/{\bf Z}_3$ orbifold
 whose six lattice vectors, or basis vectors,
 are $\beta_i$ with $i = 1,2,\cdots,6$.
The inner products of these vectors can be described by the matrix
\begin{equation}
 {\cal M} =
 \left(
 \begin{array}{cccccc}
 r_{11} & cr_{11} & r_{12} & cr_{12}-st_{12} & r_{13} & cr_{13}-st_{13} \\
 cr_{11} & r_{11} & cr_{12}+st_{12} & r_{12} & cr_{13}+st_{13} & r_{13} \\
 r_{12} & cr_{12}+st_{12} & r_{22} & cr_{22} & r_{23} & cr_{23}-st_{23} \\
 cr_{12}-st_{12} & r_{12} & cr_{22} & r_{22} & cr_{23}+st_{23} & r_{23} \\
 r_{13} & cr_{13}+st_{13} & r_{23} & cr_{23}+st_{23} & r_{33} & cr_{33} \\
 cr_{13}-st_{13} & r_{13} & cr_{23}-st_{23} & r_{23} & cr_{33} & r_{33}
 \end{array}
 \right)
\label{inner-product-M}
\end{equation}
 as
\begin{equation}
 \beta_i \cdot \beta_j = {\cal M}_{ij},
\label{inner-product}
\end{equation}
 where $c\equiv\cos(2\pi/3)=-1/2$ and $s\equiv\sin(2\pi/3)=\sqrt{3}/2$,
  and all the $r$ and $t$ moduli parameters have units of $l_s^2$
  with $l_s \equiv \sqrt{\alpha'}$.
We use the unit of $l_s=1$.
The combinations $(\beta_1, \beta_2)$, $(\beta_3, \beta_4)$ and $(\beta_5, \beta_6)$
 are SU$(3)$ lattice vectors for the three corresponding two-tori,
 and $r_{11}$, $r_{22}$ and $r_{33}$ describe their areas, respectively.
The three tori are not necessarily orthogonal with respect to one another:
 the three moduli $r_{12}$, $r_{23}$ and $r_{13}$
 can account for the lack of orthogonality of two basis vectors belonging to different tori,
 while the three other moduli $t_{12}$, $t_{23}$ and $t_{13}$
 describe ``twists'' between pairs of tori.
From the conditions
 that the $\beta_i$ are basis vectors of six-dimensional space,
 $\vert \beta_i \pm \beta_j \vert \ne 0$ for $i \ne j$,
 the possible ranges of the values of moduli parameters are restricted so that
\begin{equation}
 r_{11} > 0,
\quad
 r_{22} > 0,
\quad
 r_{33} > 0,
\end{equation}
 for what concerns the volume moduli, and
\begin{equation}
 - \frac{1}{2} \left(r_{11}+r_{22}\right)
  < r_{12} < \frac{1}{2} \left(r_{11}+r_{22}\right),
\quad
 - \frac{\sqrt{3}}{2} \left(r_{11}+r_{22}\right)
  < t_{12} < \frac{\sqrt{3}}{2} \left(r_{11}+r_{22}\right),
\end{equation}
\begin{equation}
 - \frac{1}{2} \left(r_{22}+r_{33}\right)
  < r_{23} < \frac{1}{2} \left(r_{22}+r_{33}\right),
\quad
 - \frac{\sqrt{3}}{2} \left(r_{22}+r_{33}\right)
  < t_{23} < \frac{\sqrt{3}}{2} \left(r_{22}+r_{33}\right),
\end{equation}
\begin{equation}
 - \frac{1}{2} \left(r_{11}+r_{33}\right)
  < r_{13} < \frac{1}{2} \left(r_{11}+r_{33}\right),
\quad
 - \frac{\sqrt{3}}{2} \left(r_{11}+r_{33}\right)
  < t_{13} < \frac{\sqrt{3}}{2} \left(r_{11}+r_{33}\right).
\end{equation}
These nine moduli are K\"ahler moduli,
 and there are no complex structure moduli in this orbifold,
 so that one is confronted with a situation that parallels cases where
 all complex structure moduli are stabilized by flux compactifications.
We assume, on the other hand, that dilaton is stabilized, for the sake of simplicity.

\begin{figure}[t]
\centering
\includegraphics[width=80mm]{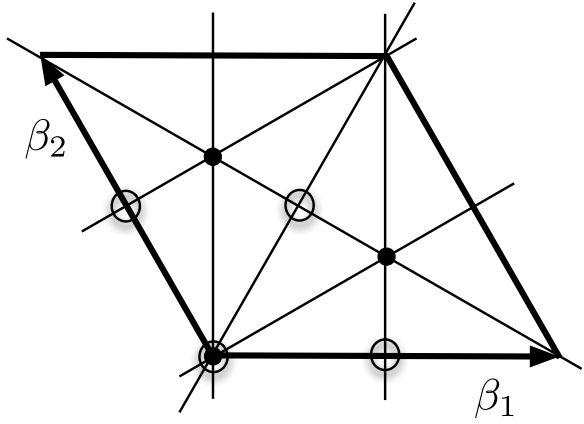}
\caption{
Fixed points in the first torus at the origin of the other four coordinates.
Three black dots indicate orbifold fixed points
 and four white circles indicate orientifold fixed points.
}
\label{fig:first-torus}
\end{figure}
Fig. \ref{fig:first-torus}
 displays points a torus spanned by $(\beta_1, \beta_2)$, the first torus,
 which correspond to the origin of the other four coordinates.
There are three orbifold fixed points inside the first torus,
 whose position vectors are
\begin{equation}
 v_0 = 0,
\quad
 v_1 = \frac{1}{3} \left( 2 \beta_1 + \beta_2 \right),
\quad
 v_2 = \frac{1}{3} \left( \beta_1 + 2 \beta_2 \right).
\end{equation}
There are four $(-1)^{F_L} R_1 R_2 R_3 \Omega$ orientifold fixed points
 inside the first torus,
 whose position vectors are
\begin{equation}
 u_0 = 0,
\quad
 u_1 = \frac{1}{2} \beta_1,
\quad
 u_2 = \frac{1}{2} \beta_2,
\quad
 u_3 = \frac{1}{2} \left( \beta_1 + \beta_2 \right).
\end{equation}
The position vectors of the fixed points in the second and third tori
 are defined in the same way as above:
\begin{equation}
 v_3 = \frac{1}{3} \left( 2 \beta_3 + \beta_4 \right),
\quad
 v_4 = \frac{1}{3} \left( \beta_3 + 2 \beta_4 \right),
\quad
 v_5 = \frac{1}{3} \left( 2 \beta_5 + \beta_6 \right),
\quad
 v_6 = \frac{1}{3} \left( \beta_5 + 2 \beta_6 \right),
\end{equation}
 for orbifold fixed points, and
\begin{equation}
 u_4 = \frac{1}{2} \beta_3,
\quad
 u_5 = \frac{1}{2} \beta_4,
\quad
 u_6 = \frac{1}{2} \left( \beta_3 + \beta_4 \right),
\quad
 u_7 = \frac{1}{2} \beta_5,
\quad
 u_8 = \frac{1}{2} \beta_6,
\quad
 u_9 = \frac{1}{2} \left( \beta_5 + \beta_6 \right),
\end{equation}
 for orientifold fixed points.
In total there are 27 orbifold fixed points, whose position vectors are
 $v_0, v_1, v_2, \cdots v_6$ and $w_1, w_2, \cdots w_{20}$ (not parallel to any single tori),
 and 64 orientifold fixed points, whose position vectors are
 $u_0, u_1, u_2, \cdots, u_{63}$.

\begin{figure}[t]
\centering
\includegraphics[width=120mm]{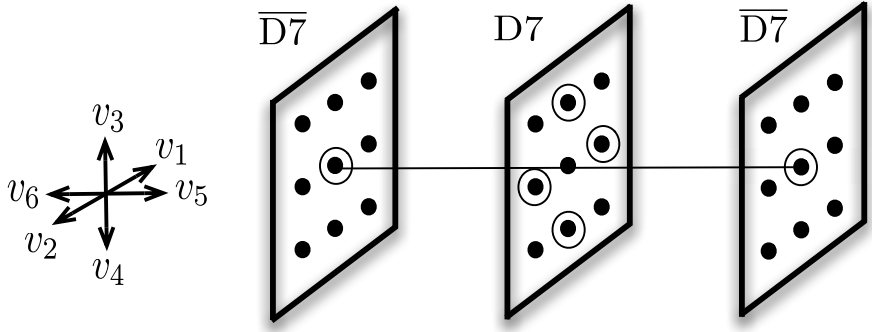}
\caption{
The D-brane configuration.
Black dots indicate orbifold fixed points.
D3-branes are placed at the positions of black dots only,
 anti-D3-branes are placed at the positions of black dots with circles.
}
\label{fig:config}
\end{figure}

Now we introduce D-branes at each fixed point and construct a concrete toy model
 in which all Ramond--Ramond tadpoles are canceled
 as in the examples proposed in \cite{Aldazabal:2000sa}.
The D-brane configuration is sketched in Fig. \ref{fig:config}.
We introduce at orbifold fixed points of $v_i$ with $i=1,2,\cdots,6$,
 anti-D3-branes with the matrix of ${\bf Z}_3$ action of
 \begin{equation}
  \gamma_\theta^{\overline{\rm D3}} =
  \left(
   \begin{array}{cc}
    \alpha {\bf 1}_2 &  \\
      & \alpha^2 {\bf 1}_2
   \end{array}
  \right)
 \end{equation}
 and at the remaining orbifold fixed points of $w_l$, with $l=1,2,\cdots,20$,
 D3-branes with the matrix of ${\bf Z}_3$ action of
 \begin{equation}
  \gamma_\theta^{{\rm D3}} = {\bf 1}_2 \ ,
 \end{equation}
 where $\alpha \equiv \exp(i 2\pi/3)$ and ${\bf 1}_n$ indicates $n \times n$ unit matrix.
At the orientifold fixed point $w_0=v_0$, we introduce
 D3-branes with the matrix of ${\bf Z}_3$ action of
 \begin{equation}
  \gamma_\theta =
  \left(
   \begin{array}{ccc}
    {\bf 1}_4 &  & \\
    & \alpha {\bf 1}_6 & \\
    & & \alpha^2 {\bf 1}_6
   \end{array}
  \right),
 \end{equation}
where the matrix of orientifold action is
 \begin{equation}
  \gamma_\Omega =
  \left(
   \begin{array}{ccc}
    {\bf 1}_4 &  & \\
    & 0 & {\bf 1}_6 \\
    & {\bf 1}_6 & 0
   \end{array}
  \right).
 \end{equation}
We also introduce D7-branes and anti-D7-branes
 that occupy the space spanned by vectors $\beta_{1,2,3,4}$ in the compact space.
We introduce at each orbifold fixed point of $v_5$ and $v_6$,
 D7-branes with the matrix of ${\bf Z}_3$ action of
 \begin{equation}
  \gamma_\theta^{\overline{\rm D7}} = {\bf 1}_6 \ ,
 \end{equation}
and at the orbifold fixed point of $v_0$,
 anti-D7-branes with the matrix of ${\bf Z}_3$ action of
 \begin{equation}
  \gamma_\theta^{{\rm D7}} =
  \left(
   \begin{array}{cc}
    \alpha {\bf 1}_6 &  \\
      & \alpha^2 {\bf 1}_6
   \end{array}
  \right),
 \end{equation}
 where the matrix of orientifold action on these anti-D7-branes is trivial.
The matrices of orbifold action indicate that
 all D-branes are fractional: they can not move away from fixed points
 respecting ${\bf Z}_3$ symmetry and $(-1)^{F_L} R_1 R_2 R_3 \Omega$ orientifold projection.
Note that, for simplicity, we assume the stabilization of twisted K\"ahler moduli
 (there are no twisted complex structure moduli in this orbifold).

The twisted Ramond--Ramond tadpoles are canceled as follows:
 at all orbifold fixed points
\begin{eqnarray}
 3 {\rm tr} \gamma_\theta^{\rm D3} + {\rm tr} \gamma_\theta^{\rm D7} = 0,
&&
 3 {\rm tr} \gamma_\theta^{\rm D3} - {\rm tr} \gamma_\theta^{\overline{\rm D7}} = 0,
\\
 -3 {\rm tr} \gamma_\theta^{\overline{\rm D3}} + {\rm tr} \gamma_\theta^{\rm D7} = 0,
&&
 -3 {\rm tr} \gamma_\theta^{\overline{\rm D3}} - {\rm tr} \gamma_\theta^{\overline{\rm D7}} = 0,
\end{eqnarray}
 are satisfied, and at an orientifold fixed point
\begin{equation}
 3 {\rm tr} \gamma_\theta + {\rm tr} \gamma_\theta^{\rm D7} = -12
\end{equation}
 is satisfied.
Since the total numbers of D3-branes and anti-D3-branes are $56$ and $24$, respectively,
 the total D3-brane Ramond--Ramond charge is $56-24=32$
 which is canceled by the D3-brane Ramond--Ramond charge of O3-plane: $(-1/2) \times 64 = -32$.
Since the total number of D7-branes, $12$, is equal to the total number of anti-D7-branes, the
 D7-brane Ramond--Ramond charge is also canceled.

On the other hand,
 since Neveu-Schwarz--Neveu-Schwarz tadpoles are not canceled,
the tensions of D-branes and O-planes are not canceled and
 the cosmological constant induced in four--dimensional low--energy effective theory is large.
This problem is common to models with supersymmetry breaking,
 and for this reason a solution relying on non-perturbative effects with three-from fluxes 
 is proposed in the KKLT scenario, for example
 \footnote{
 If the uncanceled Neveu-Schwarz--Neveu-Schwarz tadpoles remain,
 one needs to redefine the background
  \cite{Fischler,Dudas:2004nd,Kitazawa:2008hv,Pius:2014gza}.
 To be precise, the cosmological constant
  is a potential depending on the dilaton and on the other moduli,
  which could contain even their tadpoles.
  }.
Here, we do not pursue the solution to this problem in our toy model,
 because our aim is not to construct realistic models,
 but more simply to explore qualitatively possible effects of D-brane dynamics
 on moduli stabilization.

\section{Moduli potentials}
\label{potential}

The potential energy as a function of the distance between two D-branes
 can be obtained from the amplitude of closed string exchanges by Fourier transformation.
In this article
 we investigate only one closed string exchange amplitudes, while
 considering only massless modes of the closed string.
In other words, we work in low--energy supergravity approximation.

Let us first investigate the potential between two parallel D3-branes.
The only massless mode of the Neveu-Schwarz--Neveu-Schwarz sector
 which couples to a D3-brane is graviton,
 and the amplitude of graviton exchange between two parallel D3-branes is
\begin{equation}
 A_{\rm D3-D3}^{\rm NS-NS} = i V_4 {{2 \kappa_{10}^2 \tau_3^2} \over {(k_{\rm 6D})^2}},
\label{amp-NSNS}
\end{equation}
 where $V_4$ is the world volume of the D3-brane
 and $k_{\rm 6D}$ is the momentum in six-dimensional space parallel to the D3-brane.
The ten-dimensional gravitational constant and D$p$-brane tension are defined as
\begin{equation}
 2\kappa_{10}^2 \equiv (2\pi)^7 (\alpha')^4 g_s^2
\quad \mbox{and} \quad
 \tau_p \equiv \frac{1}{g_s} (2\pi)^{-p} (\alpha')^{-\frac{p+1}{2}},
\label{def-tensions}
\end{equation}
 respectively, where $g_s$ is the string coupling constant~\cite{Johnson:2003gi}.
In the presence of a compact six-dimensional space,
 we would need to consider Kaluza-Klein momentum modes,
 which would endow the potential with the proper periodicity of the compact space.
However, we do not consider Kaluza-Klein momentum modes in the analysis on our toy model,
 because they cause unnecessary complications for our aim:
 qualitatively clarifying possible universal effects of the D-brane dynamics to moduli stabilization.
We believe that
 this simplification is convenient to this aim
 and rigorous quantitative analysis on a toy model is not necessary.
One massless Ramond--Ramond field couples to a D3-brane
 and the contribution to the amplitude is
\begin{equation}
 A_{\rm D3-D3}^{\rm R-R} = - i V_4 {{2 \kappa_{10}^2 \tau_3^2} \over {(k_{\rm 6D})^2}},
\label{amp-RR}
\end{equation}
 which cancels exactly the contribution of eq.(\ref{amp-NSNS}) because of supersymmetry.
Therefore, as is well known,
 there is no potential energy, and thus no force, between parallel D3-branes.

The amplitudes between parallel anti-D3-brane, D3-brane and O3-plane
 can be obtained by considering the differences of their Ramond--Ramond charges and tensions.
\begin{eqnarray}
 A_{\rm D3-D3}
  &=& A_{\rm D3-D3}^{\rm NS-NS} + A_{\rm D3-D3}^{\rm R-R} = 0
\\
 A_{\rm \overline{D3}-D3}
  &=& A_{\rm D3-D3}^{\rm NS-NS} - A_{\rm D3-D3}^{\rm R-R}
   = i V_4 {{4 \kappa_{10}^2 \tau_3^2} \over {(k_{\rm 6D})^2}}
\\
 A_{\rm D3-O3}
  &=& - \frac{1}{2} A_{\rm D3-D3}^{\rm NS-NS} - \frac{1}{2} A_{\rm D3-D3}^{\rm R-R} = 0
\\
 A_{\rm \overline{D3}-O3}
  &=& - \frac{1}{2} A_{\rm D3-D3}^{\rm NS-NS} + \frac{1}{2} A_{\rm D3-D3}^{\rm R-R}
   = - i V_4 {{2 \kappa_{10}^2 \tau_3^2} \over {(k_{\rm 6D})^2}}
\end{eqnarray}
Therefore,
 there are non-trivial potentials between anti-D3-brane and D3-branes
 and between anti-D3-branes and O3-planes as follows:
\begin{eqnarray}
 V_{\rm \overline{D3}-D3}
  &=& - \frac{1}{iV_4} \int \frac{d^4k}{(2\pi)^6} e^{ikx} A_{\rm \overline{D3}-D3}
   = - \frac{4 \kappa_{10}^2 \tau_3^2}{4 \pi^3} \frac{1}{r^4},
\\
 V_{\rm \overline{D3}-O3}
  &=& - \frac{1}{iV_4} \int \frac{d^4k}{(2\pi)^6} e^{ikx} A_{\rm \overline{D3}-O3}
   = \frac{4 \kappa_{10}^2 \tau_3^2}{4 \pi^3} \frac{1}{2} \frac{1}{r^4},
\end{eqnarray}
 where $r$ is the distance between two parallel branes.
More precisely,
 these are potential energy densities in the four--dimensional low--energy effective theory.
We see that
 there is an attractive force between parallel anti-D3-brane and D3-brane
 and a repulsive force between parallel anti-D3-brane and O3-brane.

Let us now investigate the potential between two parallel D7-branes.
The analysis is almost the same as above.
Two massless modes of Neveu-Schwarz--Neveu-Schwarz sector, dilaton and graviton,
 can couple to a D7-brane, but only the dilaton exchange contributes.
\begin{equation}
 A_{\rm D7-D7}^{\rm NS-NS} = i V_8 {{2 \kappa_{10}^2 \tau_7^2} \over {(k_{\rm 2D})^2}},
\end{equation}
 where $V_8$ is the world volume of the D7-brane
 and $k_{\rm 2D}$ is the momentum in two dimensional space parallel to the D7-brane.
Since the contribution of the Ramond--Ramond exchange exactly cancels this amplitude,
there is a non-trivial amplitude between parallel anti-D7-brane and D7-brane
\begin{equation}
 A_{\rm \overline{D7}-D7} = i V_8 {{4 \kappa_{10}^2 \tau_7^2} \over {(k_{\rm 2D})^2}}.
\end{equation}
The corresponding non-trivial potential energy density in the low--energy four--dimensional theory
 can be obtained as \cite{Kitazawa:2014hya}
\begin{equation}
 V_{\rm \overline{D7}-D7}
  = - \frac{1}{iV_4} \int \frac{d^2k}{(2\pi)^2} e^{ikx} A_{\rm \overline{D7}-D7}
  = \frac{4 \kappa_{10}^2 \tau_3^2}{4 \pi^3}
    \frac{V_8}{2 (2 \pi)^6 V_4}
    \ln \frac{r}{\sqrt{2\pi}},
\end{equation}
 where we have used eq.(\ref{def-tensions}) with $\alpha'=1$.
In our toy model
 the ratio $V_8/V_4$ is the four--dimensional volume of D7-brane and anti-D7-branes
 in the six-dimensional compact space, which is proportional to $r_{11} \times r_{22}$.

Now we can write explicitly
 the total potential of our toy model as a function of moduli parameters.
\begin{equation}
 V(r_1,r_2,r_3,s_1,s_2,s_3,t_1,t_2,t_3)
 = V_{\rm \overline{D3}-D3-O3}^{\rm total} + V_{\rm \overline{D7}-D7}^{\rm total},
\end{equation}
\begin{equation}
 V_{\rm \overline{D3}-D3-O3}^{\rm total}
  \equiv V_{\rm \overline{D3}-D3}^{\rm total} + V_{\rm \overline{D3}-O3}^{\rm total},
\end{equation}
 where we have defined
 $r_{11}, r_{22}, r_{33}, r_{12}, r_{23}, r_{13}, t_{12}, t_{23}$ and $t_{13}$
 as $r_1, r_2, r_3, s_1, s_2, s_3, t_1, t_2$ and $t_3$, respectively.
Each contribution is given as
\begin{equation}
 V_{\rm \overline{D3}-D3}^{\rm total}
 = - \frac{4 \kappa_{10}^2 \tau_3^2}{4 \pi^3}
   \times 4 \times
   \left\{
    16 \sum_{i=1}^6 \frac{1}{( \vert v_i - w_0 \vert^2 )^2}
    + 2 \sum_{i=1}^6 \sum_{l=1}^{20} \frac{1}{( \vert v_i - w_l \vert^2 )^2}
   \right\},
\end{equation}
\begin{equation}
 V_{\rm \overline{D3}-O3}^{\rm total}
 = \frac{4 \kappa_{10}^2 \tau_3^2}{4 \pi^3}
   \times 4 \times
   \sum_{i=1}^6 \sum_{\alpha=0}^{63} \frac{1}{2} \frac{1}{( \vert v_i - u_\alpha \vert^2 )^2},
\end{equation}
\begin{equation}
 V_{\rm \overline{D7}-D7}^{\rm total}
 = \frac{4 \kappa_{10}^2 \tau_3^2}{4 \pi^3}
   \frac{V_8}{2 (2 \pi)^6 V_4}
   \times 12 \times
   \left\{
    6 \times \ln \frac{\vert v_5 \vert}{\sqrt{2\pi}}
    + 6 \times \ln \frac{\vert v_6 \vert}{\sqrt{2\pi}}
   \right\}
 = \frac{4 \kappa_{10}^2 \tau_3^2}{4 \pi^3}
   \frac{36 V_8}{(2 \pi)^6 V_4}
   \ln \frac{r_3}{6\pi}
\end{equation}
 considering number of D-branes and O3-planes at each fixed point.
Since all the position vectors of orbifold fixed points,
 $v_0, v_1, \cdots, v_6$ and $w_1, w_2, \cdots, w_{20}$,
 and those of orientifold fixed points, $u_0, u_1, \cdots, u_{63}$,
 are described by six basis vectors $\beta_1, \beta_2, \cdots, \beta_6$
 with inner product defined in eq.(\ref{inner-product}),
 the potential is an explicit function of nine moduli parameters.
Since we do not account for the torus periodicity, for the sake of convenience,
 the potential is the function of shortest distances between every two objects.
There is an exact symmetry
\begin{equation}
 V(r_1,r_2,r_3,s_1,s_2,s_3,t_1,t_2,t_3)
 = V(r_1,r_2,r_3,s_1,s_2,s_3,-t_1,-t_2,-t_3),
\label{exact-sym}
\end{equation}
 since the configuration of D-branes and O-planes is symmetric under the replacement
\begin{equation}
 (\beta_1,\beta_2) \rightarrow (\beta_2,\beta_1),
\quad
 (\beta_3,\beta_4) \rightarrow (\beta_4,\beta_3)
\quad \mbox{and} \quad
 (\beta_5,\beta_6) \rightarrow (\beta_6,\beta_5),
\end{equation}
 and this replacement flips the signs in front of $t_{12}, t_{23}$ and $t_{13}$
 in the matrix of eq.(\ref{inner-product-M}).

Let us first analyze the potential $V_{\rm \overline{D3}-D3-O3}^{\rm total}$
 with the symmetric ansatz
\begin{equation}
 r_1 = r_2 = r_3 = r,
\quad
 s_1 = s_2 = s_3 = s
\quad \mbox{and} \quad
 t_1 = t_2 = t_3 = t
\end{equation}
in order to elicit its overall nature.
\begin{figure}[t]
\centering
\includegraphics[width=50mm]{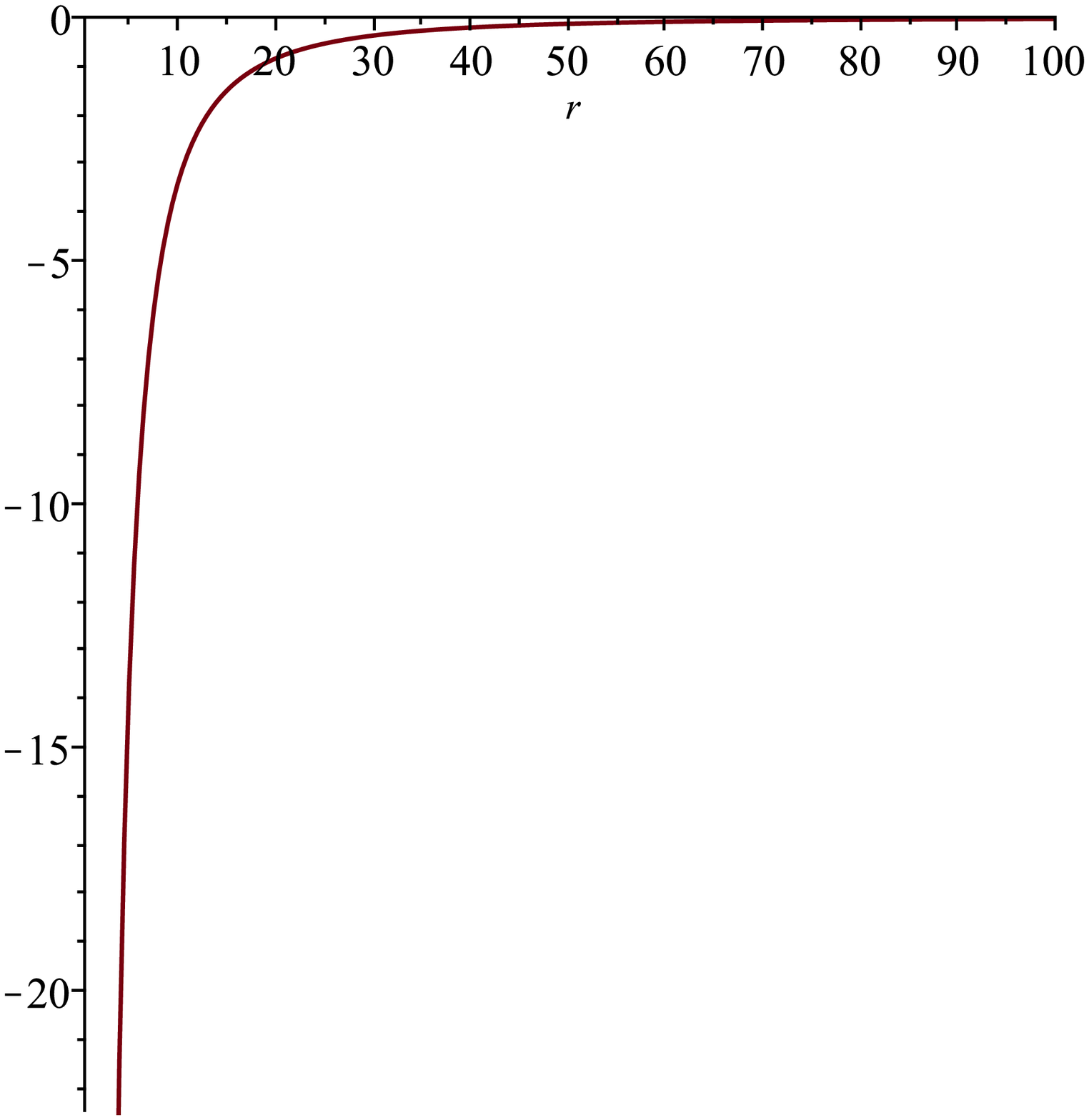}
\includegraphics[width=50mm]{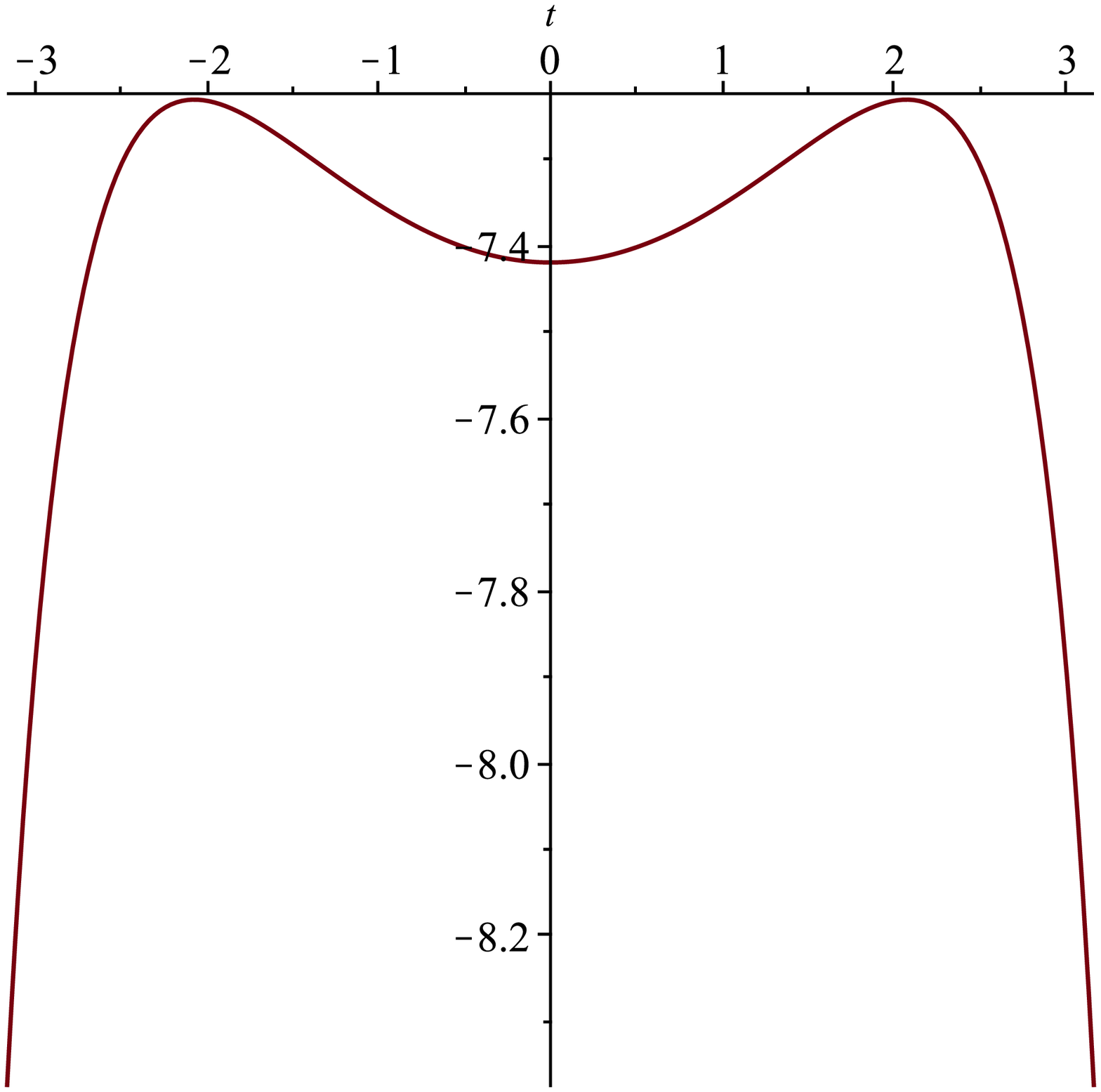}
\includegraphics[width=50mm]{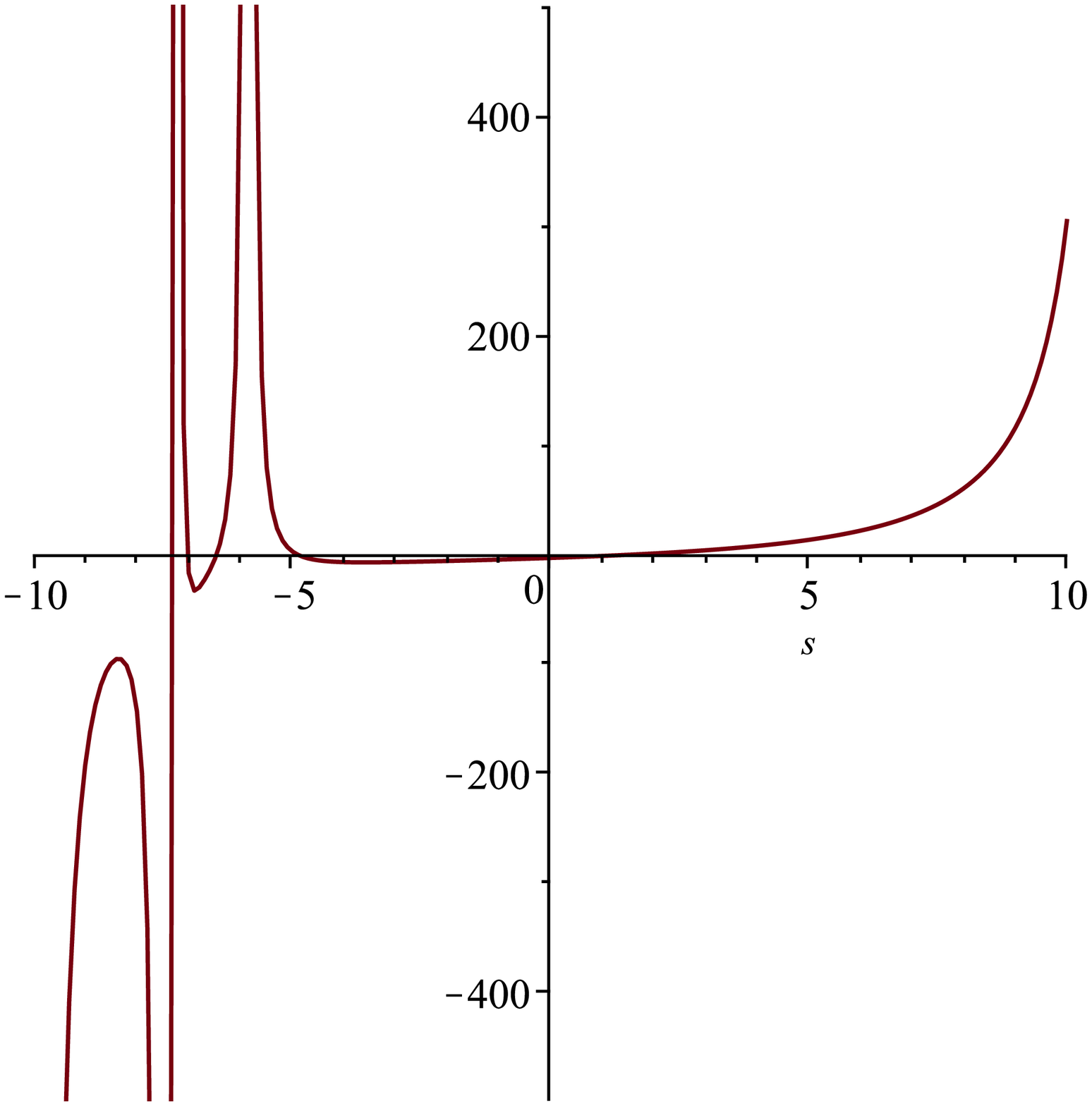}
\caption{
The form fo the potential in the symmetric ansatz.
From left to right:
 the plot of $V_{\rm \overline{D3}-D3-O3}^{\rm total}(r,0,0)$,
 the plot of $V_{\rm \overline{D3}-D3-O3}^{\rm total}(10,-3.6,t)$
 and
 the plot of $V_{\rm \overline{D3}-D3-O3}^{\rm total}(10,s,0)$
 which has local minimum at $s \simeq -3.6$.
}
\label{fig:symmetric}
\end{figure}
The form of the function $V_{\rm \overline{D3}-D3-O3}^{\rm total}(r,0,0)$
 in the left plot of Fig. \ref{fig:symmetric}
 represents an attractive force.
One can see that the total attractive force between anti-D3-branes and D3-branes
 overcomes the total repulsive force between anti-D3-branes and O3-planes.
The volume moduli $r$ is not stabilized by the D-brane dynamics only.
In other words, volume moduli stabilization
 should be achieved by other mechanisms
 under this kind of unstable potential that has replaced the original flat potential.
The exact symmetry
\begin{equation}
 V_{\rm \overline{D3}-D3-O3}^{\rm total}(r,s,t)
  = V_{\rm \overline{D3}-D3-O3}^{\rm total}(r,s,-t)
\end{equation}
 indicates that $t=0$ is a stationary point.
It is actually a local minimum of the potential with a finite value of $r$ and $s$,
 as can be seen in middle plot in Fig. \ref{fig:symmetric}.
This suggests the stabilization of $t_{12}, t_{23}$ and $t_{13}$ by the D-brane dynamics.
At several values of $s$ (non-orthogonality of two-tori) the potential diverges,
 as can seen from the right plot in Fig. \ref{fig:symmetric},
 because the position of some orbifold fixed point
 coincides with the position of other orbifold fixed point or orientifold fixed point.
For example,
 the distance between the orbifold fixed points $v_2$ and $w_5 = v_1 + v_3 + v_5$
\begin{equation}
 \vert v_2 - w_5 \vert^2
 = \frac{1}{3} (r_{11} + r_{22} + r_{33})
 + \frac{1}{3} (r_{12} + 2 r_{23} + r_{13})
 - \frac{1}{\sqrt{3}} (t_{12} + t_{13})
\end{equation}
 vanishes at $s= -(3/4)r > -r$ with $t=0$, which yields a negative divergence
 resulting from the coincident position of anti-D3-branes and D3-branes.
The distance between the orbifold fixed point $v_2$
 and the orientifold fixed point $u_{37}=(\beta_1+\beta_3+\beta_5)/2$
\begin{equation}
 \vert v_2 - u_{37} \vert^2
 = \frac{7}{12} r_{11} + \frac{1}{4} (r_{22} + r_{33})
 + \frac{1}{2} (r_{12} + r_{23} + r_{13})
 - \frac{1}{\sqrt{3}} (t_{12} + t_{13})
\end{equation}
 vanishes at $s= -(13/18)r > -r$ with $t=0$, which gives rise to a positive divergence
 resulting from coincident anti-D3-branes and O3-plane.
Around such values of $s$ (or $r_{12}, r_{23}$ and $r_{13}$)
 the distances between D-branes and O-planes are shorter than $l_s$,
 and our analysis based on closed string exchanges and low--energy supergravity approximation
 is not valid.
The conclusion is that one should be careful in a general analysis without symmetric ansatz.

Let us next discuss the situation where
 volume moduli are fixed as $r_{11}=r_{22}=r_{33}=r_{\rm fixed} > 0$
 by some mechanism, by exponential potentials from non-perturbative effects
 as in the KKLT scenario, for example.
We can find a stationary point of the potential $V_{\rm \overline{D3}-D3-O3}^{\rm total}$
 at $t_1=t_2=t_3=0$ and a negative value of $s_1=s_2=s_3$
 depending on the value of $r_{\rm fixed}$
 (a numerical analysis gives $s_1=s_2=s_3\simeq-3.6$ for $r_{\rm fixed}=10$).
At this point the directions
\begin{equation}
 \frac{1}{\sqrt{3}} (t_1+t_2+t_3),
\quad
 \frac{1}{\sqrt{2}} (t_1-t_2),
\quad
 \frac{1}{\sqrt{6}} (t_1+t_2-2t_3)
\quad \mbox{and} \quad
 \frac{1}{\sqrt{3}} (s_1+s_2+s_3)
\end{equation}
 are stabilized, but the directions
\begin{equation}
 \frac{1}{\sqrt{2}} (s_1-s_2)
 \quad \mbox{and} \quad
 \frac{1}{\sqrt{6}} (s_1+s_2-2s_3)
\end{equation}
 are not stabilized.
This is an example of partial K\"ahler moduli stabilization by D-brane dynamics.
There are many other stationary points,
 but there is no point at which all the remaining moduli are stabilized.

Finally, let us consider the effect of $V_{\rm \overline{D7}-D7}^{\rm total}$
 which gives an attractive force inside the third torus.
Let us assume the stabilization of $r_1=r_2=r_{\rm fixed}$ and $s_1=s_2=s_3=0$ by some mechanisms,
 and let us investigate the stabilization of $r_3$.
The potential
\begin{equation}
 V_{\rm \overline{D3}-D3-O3}^{\rm total}(r_{\rm fixed},r_{\rm fixed},r_3,0,0,0,0,0,0)
\end{equation}
 displays an interesting behavior.
For short distances $r_3 < r_{\rm fixed}$ the potential gives an attractive force,
 but on the other hand for larger distances $r_3 > r_{\rm fixed}$
 the potential gives rise to a repulsive force.
The way of balancing attractive forces between anti-D3-branes and D3-branes
 and repulsive forces between anti-D3-branes and O3-planes
 is different from what we saw with the symmetric ansatz.
The long distance force
 is essentially repulsive under the condition that two volume moduli are fixed.
Since the number of terms of the form $1/(r_{\rm fixed} - r_3)^2$
 in the potential from the combinations of anti-D3-brane and O3-plane
 is larger than that arising from combinations of anti-D3-brane and O3-planes,
 repulsive forces are more extensively ``cut off'' at shorter distances than attractive ones.
Therefore,
 the stabilization of $r_3$ is possible
balancing the repulsive force
 from the power potential at long distance, $r_3 > r_{\rm fixed}$,
 and the attractive force by the logarithmic potential.
We set $V_8/V_4 = r_1 \times r_2 = r_{\rm fixed}^2$ in $V_{\rm \overline{D7}-D7}^{\rm total}$, and
 therefore larger values of $r_{\rm fixed}$ give a stronger attractive force
 from $V_{\rm \overline{D7}-D7}^{\rm total}$.
On the other hand,
 since larger values of $r_{\rm fixed}$
 give a wider $r_3$ range of the attractive force from $V_{\rm \overline{D3}-D3-O3}^{\rm total}$,
 there may be some appropriate values of $r_{\rm fixed}$ for stabilization.
A numerical analysis shows that stabilization occurs at $r_3 \simeq 110$ with $r_{\rm fixed}=25$.
This is another example of partial K\"ahler moduli stabilization by D-brane dynamics.

\section{Conclusions}
\label{conclusions}

We have investigated
 the effect of the mutual forces among D-branes, anti-D-branes and O-planes on moduli stabilization
 in a concrete, rich and yet relatively simple example.
Although our analysis has relied on a specific reference toy model,
 we believe to have captured some general features of the effect.
These types of setting are of interest in the context of supersymmetry breaking,
 and have the virtue of exhibiting, in concrete examples,
 the resulting back reaction to the moduli stabilization.
As we have seen, the moduli potential typically suffers from some instabilities,
 which ought to be overcome by some other mechanism if the models at stake are to acceptable.
If this were not the case, in the KKLT scenario, for example,
 the lifetime of de Sitter vacuum could become unphysically short in some cases.
On the other hand,
 it is possible that D-brane interactions have a beneficial effect on moduli stabilization,
 via the balance of forces.
In addition,
 the extremal value of the potential energy density at the stationary point can be negative,
 thus providing a negative contribution to the cosmological constant
 in the four--dimensional low--energy effective theory,
 in addition to those arising from orientifolds
 or from non-perturbative effects associated to three-from fluxes.
We note that it can not be expected in general
 that all the K\"ahler moduli are stabilized only by D-brane dynamics,
 because of no stable equilibrium for freely moving charges,
 although possible motions of fractional D-branes and O-planes are restricted.

We have neglected Kaluza-Klein modes of closed string propagation in constructing the potentials.
Taking into account the compact nature of the internal space
 would make the potentials very complicated.
The distance dependence of the force
 should be modified by additional contributions from infinite images along the straight lines
 joining the two sources, which could favor or disfavor moduli stabilization.
The effects which have discussed in this article
 could be amplified or suppressed, but would never be disappeared.
A detailed quantitative analysis along these lines, however,
 would be appropriate and necessary for more realistic models.
We have also neglected another possible type of back reaction,
 modifications of the background geometry
 including the blowing-up of orbifold singularities
 due to possible Neveu-Schwarz--Neveu-Schwarz tadpoles, as well as the
 destabilization of twisted K\"ahler moduli.
These changes would affect the definition of distance,
 and would clearly have a bearing on the argument of force balance,
 but unfortunately our present techniques do not suffice to treat this problem.
On the other hand, small steps like those presented here
 are instrumental to identify non-trivial string models without supersymmetry,
 above and beyond their perturbative definition via the ten-dimensional theory.

\section*{Acknowledgments}

The author would like to thank
 Augusto Sagnotti for valuable comments and careful reading of the manuscript.

\end{document}